\documentstyle[epsf,prl,aps,multicol]{revtex}
\begin{document}
\title{Thermodynamic and Tunneling Density of States of the Integer 
Quantum Hall Critical State}
\author{S.-R. Eric Yang$^1$, Ziqiang Wang$^2$, and A.H.MacDonald$^{3,4}$}
\address{
1)Department of Physics, Korea University, Seoul 136-701, Korea \\
2)Department of Physics, Boston College, Chestnut Hill, MA 02167 \\
3)Department of Physics, Indiana University, Bloomington, Indiana 47405 \\
4)Department of Physics, University of Texas at Austin, Austin TX 78712 }
\maketitle
\draft
\begin{abstract}
We examine the long wave length limit of the self-consistent 
Hartree-Fock approximation
irreducible static density-density response function by evaluating the 
charge induced by an external charge.  
Our results are consistent with the compressibility sum
rule and inconsistent with earlier work that did not account for 
consistency between the exchange-local-field and the disorder potential. 
We conclude that the thermodynamic density of states is finite,
in spite of the vanishing tunneling density of states 
at the critical energy of the integer quantum Hall transition.
\end{abstract}

\thispagestyle{empty}
\pacs{PACS numbers: 73.20.Dx, 73.20.Mf}
\begin{multicols}{2}

A two dimensional(2D) electron gas in the strong magnetic field limit
has an zero-temperature insulator-metal-insulator transition 
in the presence of random disorder.
When the strength of the magnetic field sweeps past a critical value,
the off-diagonal (Hall) conductance changes abruptly by $\frac{e^2}{h}$ while 
the diagonal (dissipative) conductance rises from zero to a finite value
of order $e^2/2h$ and then returns to zero.  
This transition is often referred to as the integer quantum Hall transition.
The filling factor range of the intermediate metallic phase is 
believed to collapse to a conducting critical point in the low temperature thermodynamic limit.
According to the non-interacting theory of the integer quantum Hall effect, this behavior 
occurs because single-electron 
states are extended at an isolated critical energy near the middle of 
each disorder broadened Landau level.
The insulator-metal-insulator transition takes place 
when the Fermi level of the 2D electron system and 
one of the discrete critical energies cross.

Our current understanding of the integer quantum
Hall transition \cite{Kli} is far from complete, however, since  
Coulomb interaction between electrons are expected to provide a relevant perturbation
at the non-interacting fixed point \cite{Lee}.
In addition there are several discrepancies between the results of the 
non-interacting theory
and the experimental findings which may be due to the role of interactions.  
For example, recent experimental work
has shown that the tunneling density-of-states (DOS) 
vanishes linearly at the Fermi 
energy \cite{Chan} in sharp contrast to the finite DOS of the
non-interacting theory.
To account for this discrepancy Yang and MacDonald \cite{Yang1}
carried out a numerical study in which disorder was treated exactly and 
Coulomb interaction 
was described by a self-consistent Hartree-Fock (HF) approximation.  
They found a linear 
Coulomb gap at all filling factors of the lowest Landau level, 
even at the critical energy.
Wang and Xiong\cite{Wang} investigated the effects of  
dynamical screening in a systematic nonperturbative
resummation of the most singular diagrams.
They found that the DOS is linear 
at the critical energy, as in the Hartree-Fock theory, 
exhibiting quantum Coulomb gap behavior.
In spite of the qualitative DOS change due to interactions,
however, Yang et al.\cite{Yang2}
found that the value of the localization
length exponent $\nu_{loc}$ is identical to that of the 
non interacting theory \cite{Aoki,Cha,Huo,Koch,Sha}. 

A second discrepancy between non-interacting electron theory 
and experimental findings 
is the value of the dynamical scaling
exponent, $z$.  For non-interacting electrons,  
$z$ equals the space dimension $d=2$. This remains true for 
short-ranged interactions since they are irrelevant at the
non-interacting fixed point \cite{Lee,Wangetal}.
However, experimental data are consistent with $z\approx 1$ \cite{Eng}.
Lee and Wang \cite{Lee} conjectured that a change in $z$ 
with no change in $\nu_{loc}$ 
is due to the non-critical linear suppression of the single-particle 
DOS induced by Coulomb
interactions.\cite{Yang1,Yang2}.  
Recently, Huckestein and Backhaus\cite{Huc} attempted to 
substantiate this conjecture by evaluating the density-density 
response function in the 
quantum Hall regime, including interaction effects within a time 
dependent Hartree-Fock
approximation (TDHFA).  
Their analysis of dynamic scaling gives $z=1$ and appears to be 
consistent with the result of a naive
scaling argument incorporating the linearly vanishing DOS: the 
frequency-dependent length scale  
$L_{\omega}=\omega^{-1/z}=\frac{1}{\sqrt{\rho_0\hbar\omega}}$
with density-of-states $\rho_0 \propto \omega$ implies $z=1$.
However, it is not clear that  the $\rho_0$ that enters $L_\omega$
is the tunneling DOS.  In fact, it has been argued \cite{Plee} 
that the relevant DOS here is
the thermodynamics DOS or the compressibility $dn/d\mu$, which 
should be smooth and finite for a
disordered system on general grounds.   
Huckestein and Backhaus, however,  found that the thermodynamic 
DOS or the compressibility is zero
at the transition.  Since the compressibility is
proportional to the inverse screening length, 
$q_{sc}=2\pi e^2\frac{dn}{d\mu}$, 
their result would imply an infinite screening length.  
The screening properties of the conducting
critical state would therefore be 
highly unconventional.
This result is plausible since 
the single particle states at a quantum Hall transition
have the special property that 
only states at the critical energy are delocalized 
while all other states are localized. 
However, in order to make their calculation numerically manageable,
Huckestein and Backhaus\cite{Huc}
were forced to neglect correlations between the Hartree-Fock 
theory self-energy and 
the random potentials.  Closely related approximations have long been 
common in a number of circumstances,
for example in addressing the properties of disordered superconductors
where it has recently
been demonstrated that they are unreliable.\cite{atkinson}.

In order to evaluate the Hartree-Fock irreducible density-density 
response function directly, it 
is necessary to solve a complicated integral equation.
Huckestein and Backhaus circumvented this difficulty by disorder-averaging 
the quasi-particle response 
function and the self-energy separately. 
We circumvent this difficulty without making approximations,  by 
evaluating only the total charge response $Q$ to the external potential 
from a point charge, 
solving the Hartree-Fock equations for the self-energy that is 
consistent\cite{com} with each disorder realization. 
In linear response,
$Q = \lim_{q \to 0} \chi(q) 2 \pi e^2/\epsilon q$ 
where 
$\chi(q)$ is the full static response function, related to the 
irreducible response by
$\chi(q) = \chi^{irr}(q)/[1 + (2 \pi e^2/\epsilon  q) \chi^{irr}(q)]$,
and $\epsilon$ is the dielectric constant. It follows that 
\begin{equation}
Q^{-1} = 1+ \lim_{q \to 0} \frac{\epsilon q}{2 \pi e^2 \chi^{irr}(q)}.
\label{charge}
\end{equation} 
In the approximation studied by Huckestein and Backhaus\cite{Huc},
it was determined numerically (see their Fig.~3 and discussions) that
$\chi_{HB}^{irr}(q)=s{\epsilon q\over e^2}+{\cal O}(\ell/L)$,
where $s\simeq0.2$ is a numerical constant and $\ell$ is the magnetic length.
This would imply that the second 
term on the right hand side of Eq.~(\ref{charge}) has a finite value
and $Q_{HB}\simeq0.56$. On the other hand, it is normally the case for metals
that the screening wave-vector
$q_{sc}=2\pi e^2\lim_{q \to 0} \chi^{irr}(q)$, is related
to the thermodynamic density-of-states according to
$q_{sc}=2\pi e^2 dn/d\mu$ \cite{note}, 
a result often referred to as the compressibility sum rule, which
would imply that $Q=1$.  
(It is clear from experiment that $dn/d\mu$ must be finite at the 
integer quantum Hall transition.) 
Our numerical results support the applicability of the compressibility 
sum rule and are clearly inconsistent with the value of $Q_{HB}$ that
violates this sum rule by about $50\%$. When this rule is satisfied $Q=1$, 
the external charge is perfectly screened,
and the screening length $q_{sc}$ is finite.
Our results show that despite the linearly vanishing Coulomb gap in 
the tunneling DOS, $\chi^{irr}(q=0)$ is finite and imply that 
transport is indeed governed by diffusion, as for non-interacting electrons.

We perform our calculations in the Landau gauge $[\vec{A}=(0,Bx,0)]$ and
apply quasi-periodic boundary conditions to the HF single-particle 
orbitals inside a square with area $A =a^2$.  The basis states used to 
represent the HF Hamiltonian are related to elliptic theta functions 
and can be labeled by a set of guiding centers $|X_j>$, $j=1,...,N_{\phi}$ 
inside the fundamental cell of the finite system 
($N_{\phi}$ is the total flux quantum passing through the area A).  
The wave functions of the basis states are 
$<\vec{r}|X_j>=\sum_{k=-\infty}^{\infty}\phi_{X_j+ka}(\vec{r})$
where 
\begin{eqnarray}
\phi_{X}(\vec{r})=\sqrt{ \frac{1}{a\sqrt{\pi}\ell}}\exp(i\frac{Xy}{\ell^2}
-\frac{(X-x)^2}{2\ell^2} ).
\end{eqnarray}
The HF eigenstate is given by a linear combination of the basis states. 
$|\alpha>=\sum_i |X_i><X_i| \alpha>$.
The coefficients $<X_i| \alpha>$ satisfy a set of matrix equations for the
HF Hamiltonian, $H_{HF}=V_{HF}+V_I+V_{ion}$, which is a sum of the HF potential 
in the LLL, the impurity potential,
and the ionic potential of the inserted test charge:
\begin{eqnarray}
&&\sum_i[<X_j|V_I|X_i>+<X_{j'}|V_{ion}|X_j>\nonumber\\
&&+<X_j|V_{HF}|X_i>]<X_i|\alpha>=  \epsilon_{\alpha}<X_j|\alpha>.
\end{eqnarray}
The position of the ion is $\vec{R}=a(\frac{1}{2},\frac{1}{2})$.
The matrix elements of the ion potential is
\begin{eqnarray}
&&<X_{j'}|V_{ion}|X_j>=     -\frac{1}{A}\sum_{\vec{q}}\frac{2\pi e^2}{\epsilon q}  
e^{\vec{q}\cdot\vec{R}}\nonumber\\
%&&-\frac{1}{A}\sum_{\vec{q}}\frac{2\pi e^2}{\epsilon q}
%\frac{1}{M}\sum_{k,k'=-\infty}^{\infty}<X_{j'}+k'a|e^{i\vec{q}
%\vec{r}}|X_j+ka>=\nonumber\\
&&\times \delta'_{j',j+m_y}
\exp(iq_x(X_j+q_y\ell^2/2)-q^2\ell^2/4),
\end{eqnarray}
where $\vec{q}=\frac{2\pi}{a}(m_x+sN_{\phi},m_y+tN_{\phi})$ 
with $m_{x,y}=1,...,N_{\phi}$ and $s,t$ are integers.
We have used a model disorder potential consisting 
of $N_I$ delta-function scatters
with strength uniformly distributed between $-\lambda$ 
and $\lambda$ and scattering centers 
uniformly distributed inside the fundamental 
cell of the finite-size system.
Hartree and Fock potentials can be expressed as a function
of the electron density
\begin{eqnarray}
<X_j|V_{HF}|X_i>&=&\sum_{\vec{q}}\Delta(\vec{q})
\exp[\frac{iq_x(X_i+X_j)}{2}]\nonumber\\
&&\times \delta'(j,i+m_y)U_{HF}(\vec{q}).
\end{eqnarray}
Here the sum over $\vec{q}$ is over the discrete set of wave vectors 
consistent with the boundary conditions and $\delta'(n,n')$ is $1$ 
if $n'=n ({\rm mod}N_{\phi})$ and $0$ otherwise.
$U_{HF}(\vec{q})$ is proportional to $e^2/\epsilon\ell$ and 
includes both Coulomb and exchange interactions. 
The quantity $\Delta(\vec{q})$ is proportional to the Fourier component of 
the charge density and is calculated self-consistently from the 
eigenvectors of the HF Hamiltonian by
\begin{eqnarray}
&&\Delta(\vec{q})=\frac{1}{N_{\phi}}\sum_{j=1}^{N_{\phi}} 
   \sum_{j'=1}^{N_{\phi}} \delta'(j,i+m_y)\nonumber\\
&& \times exp[\frac{iq_x(X_i+X_j)}{2}]
\sum_{\alpha=1}^N<X_{j'}|\alpha><\alpha|X_j>,
\end{eqnarray}
where $N$ is the number of electrons in the system.
For each disorder realization the matrix equation is solved for several values of 
$\gamma=(e^2/\epsilon \ell)/\Gamma$ where 
$\Gamma=( \lambda^{2}N_{I}/\ell^{2}A )^{1/2}$ is
the characteristic disorder potential energy scale and $N_I$ is the
number of impurities.
For this model the self-consistent
Born approximation (SCBA) DOS 
is nonzero for $|\epsilon|<(2/3\pi)^{(1/2)}\Gamma\sim 0.46\Gamma$.

For a system of N electrons in the $k$-th realization of the random disorder 
potential, we write the electron density as 
$\rho_{P,N}^{(k)}(\vec{r})=\sum_{\alpha=1,...,N}|\Psi_{\alpha}^{(k)}(\vec{r})|^2$
where $P=0,I$, depending on whether the ion is absent or present.
Let $\rho_{0,N}$ be the density of the uniform background charge.
When $\rho_{0,N}^{(k)}(\vec{r})$ is disorder-averaged over $N_D$ number of
realizations we find
$\overline{\rho_{0,N}(\vec{r})}\equiv 
\frac{1}{N_D}\sum_k \rho_{0,N}^{(k)}(\vec{r})=\rho_{0,N}$.
The charged density induced by the ion may be evaluated by 
computing $\overline{\rho_{I,N+1}(\vec{r})}-\overline{\rho_{0,N}(\vec{r})}=
\overline{\rho_{I,N+1}(\vec{r})}-\rho_{0,N}$.
Since 
$\overline{\rho_{0,N}(\vec{r})}=\overline{\rho_{0,N+1}(\vec{r})}-\frac{1}{A}$
the induced density $\overline{\Delta \rho(\vec{r})}=
\overline{\rho_{I,N+1}(\vec{r})}-\overline{\rho_{0,N+1}(\vec{r})}+\frac{1}{A} $.

Figure 1 displays the charge induced when an 
ion is placed at the center of the square at LLL filling
fraction $\nu=1/2$ and
for the  parameters $\gamma=0.6$, $N_D=4$
and $N_{\phi}=200$. 
% ZWSRY: Check carefully - details will depend on final results 
The inset in Fig.~1 shows how the total induced charge changes 
as the distance from the ion is varied.  These results are 
consistent with $Q=1$ and a finite screening length $\lambda_{sc} \simeq  
5\ell$. They are clearly inconsistent with Huckestein and Backhaus's result 
$Q_{HF}\simeq 0.56$. Using the DOS of the SCBA, we obtain a shorter
Thomas-Fermi screening length, 
$\lambda_{TF} = 2\pi/( \gamma \epsilon) \simeq 1 \ell$.  
Figure 2 shows the disorder averaged $\overline{\Delta\rho(x,y_i)}$   
for a much broader Landau level with $\gamma=0.1$,
$\nu=1/2$, $N_D=10$,
and $N_{\phi}=162$.  In this case  
we find that the screening length $\lambda_{sc} \simeq 11\ell$ while 
$\lambda_{TF} \simeq 6\ell$.
Our results for the $\gamma$-dependence of the screening length 
is in qualitative agreement with that implied by the DOS behavior in the
SCBA. For a more quantitative comparison  for the disorder dependence
of the screening length, a Hartree non-linear screening treatment, 
in which disorder is treated in the SCBA \cite{Yang3}, 
may be required instead of the simple SCBA.

We have also investigated the screening properties 
in the insulating regime at  $\nu=1/7$.
Figure 3 displays  the induced densities for 
$\gamma=0.1$, $N_D=10$, 
and $N_{\phi}=175$.    
We observe from the plot that the induced density does not decay rapidly and that
the fluctuations are spread over a wide range.
In this case it is difficult to define 
a screening length.  We believe this reflects
the incomplete screening property\cite{com1} of an insulator.  Indeed from the inset we see that the screening
length appears to diverge with the finite system size.  However, in the presence
of a weaker random disorder with $\gamma=0.6$, where the tendency towards crystallization is strong,
we find a different result.
Contour plots of $\rho_0^{(k)}(\vec{r})$ and $\rho_I^{(k)}(\vec{r})$  
are shown in Fig.4.
Plot (a) is without the ion while plot (b) is with the ion.
Comparing these two plots we can infer that the ion captures an electron 
while the location of the other electrons are unchanged.  
A similar abrupt reconfiguration of the ground state
was found and studied in the pinning of Wigner crystals \cite{ruzin}.

We have shown here that despite the linearly vanishing Coulomb gap in the 
tunneling DOS, the thermodynamic DOS of the integer quantum Hall
critical state remains finite, and transport is indeed governed
by diffusion as for non-interacting electrons. 
It still remains an open question whether an improved implementation of the
TDHFA, that takes into account self-consistently self-energy
and vertex corrections, can give rise to the observed dynamic critical exponent for
the integer quantum Hall transition.

SREY is supported in part by Grant No.2000-2-12-001-5 
from the interdisciplinary 
Research program of the KOSEF.  AHM is supported 
by the Welch foundation and by the National Science Foundation under grant
DMR0115947.  ZW is supported by DOE Grant No. DE-FG02-99ER45747 
and by the Research Corporation.

\begin{figure}
%\hspace{1truecm}
%\vspace{-0.5truecm}
\center
\centerline{\epsfysize=2.3in
\epsfbox{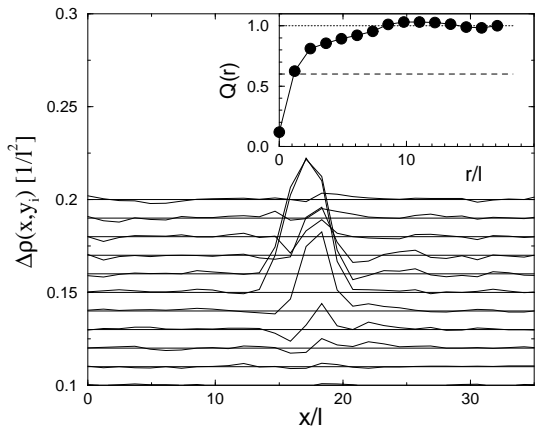}}
%\vspace{-0.5truecm}
\begin{minipage}[t]{8.1cm}
\caption{Induced charge distribution for the critical state.
Each curve represents one of  $\overline{\Delta \rho(x,y_i)}$
for $i=10,11,...,19,20$, where $y_i=\frac{a(i-1)}{29}$.  These curves
are shifted vertically for the sake of clear presentation.
The parameters are $\nu=1/2$, $\gamma=0.6$, $N_D=4$,
and $N_{\phi}=200$.  The unit
of length is $\ell$.
Inset: Integrated induced charge as a function of the distance to the
position of the ion. The dashed line corresponds to $Q_{HB}\simeq0.56$.}
\label{figure1}
\end{minipage}
\end{figure}

\begin{figure}
%\hspace{1truecm}
%\vspace{-0.5truecm}
\center
\centerline{\epsfysize=2.3in
\epsfbox{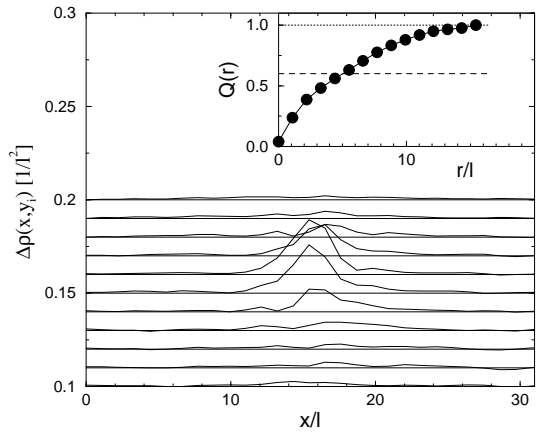}}
%\vspace{-0.5truecm}
\begin{minipage}[t]{8.1cm}
\caption{
{Same as in Fig.1 but for a more disorder broadened LLL with $\gamma=0.1$, $N_D=10$, and $N_{\phi}=162$.}
}
\label{figure2}
\end{minipage}
\end{figure}

\begin{figure}
%\hspace{1truecm}
%\vspace{-0.5truecm}
\center
\centerline{\epsfysize=2.3in
\epsfbox{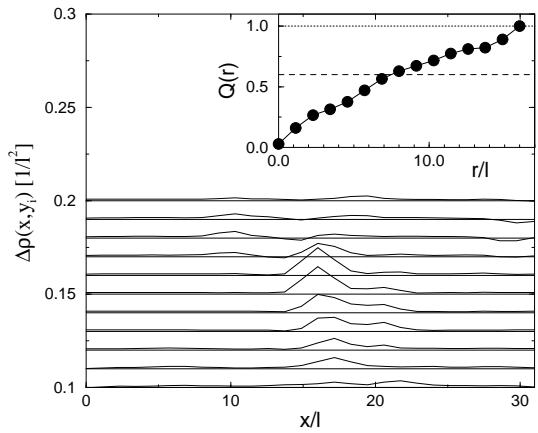}}
%\vspace{-0.5truecm}
\begin{minipage}[t]{8.1cm}
\caption{
Same as in Fig.1 but for an insulating state with
$\nu=1/7$, $\gamma=0.1$, $N_D=10$,
and $N_{\phi}=175$.}
\label{figure3}
\end{minipage}
\end{figure}

\begin{figure}
%\hspace{1truecm}
%\vspace{-0.5truecm}
\center
\centerline{\epsfysize=2.0in
\epsfbox{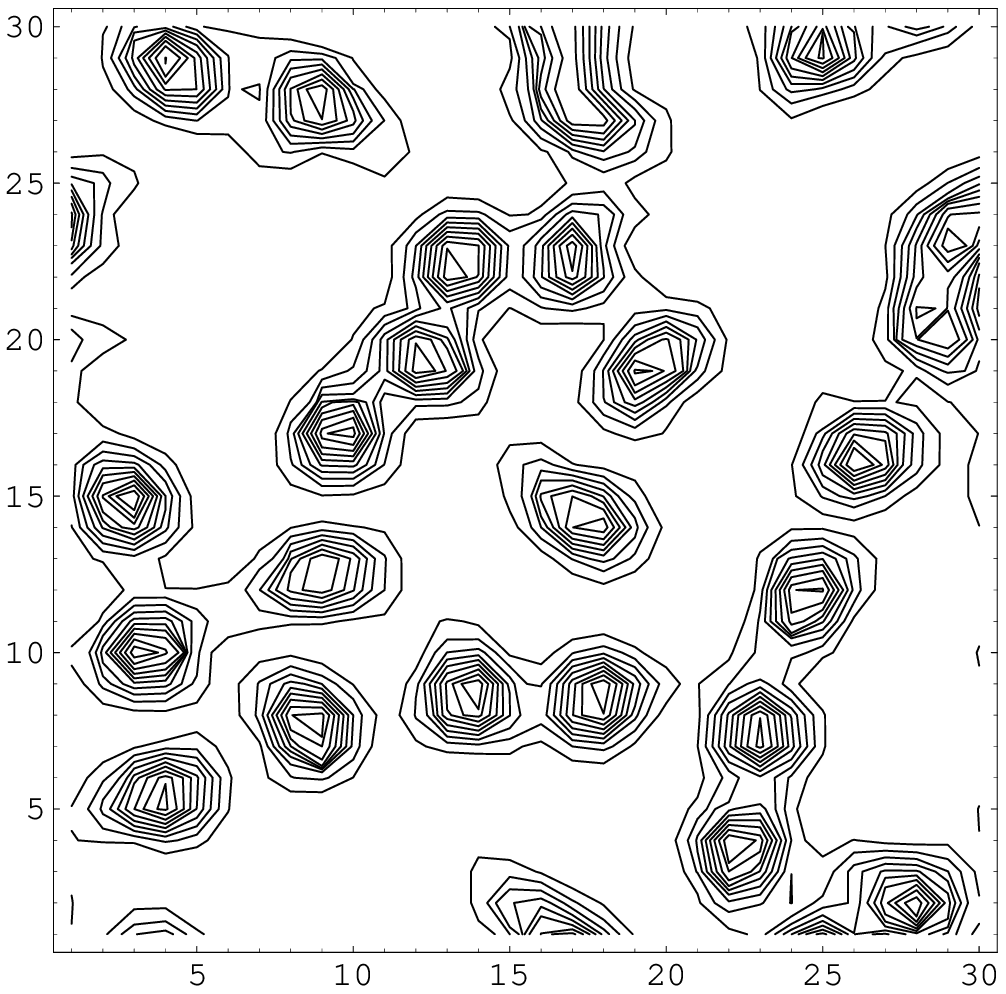}}
\centerline{\epsfysize=2.0in
\epsfbox{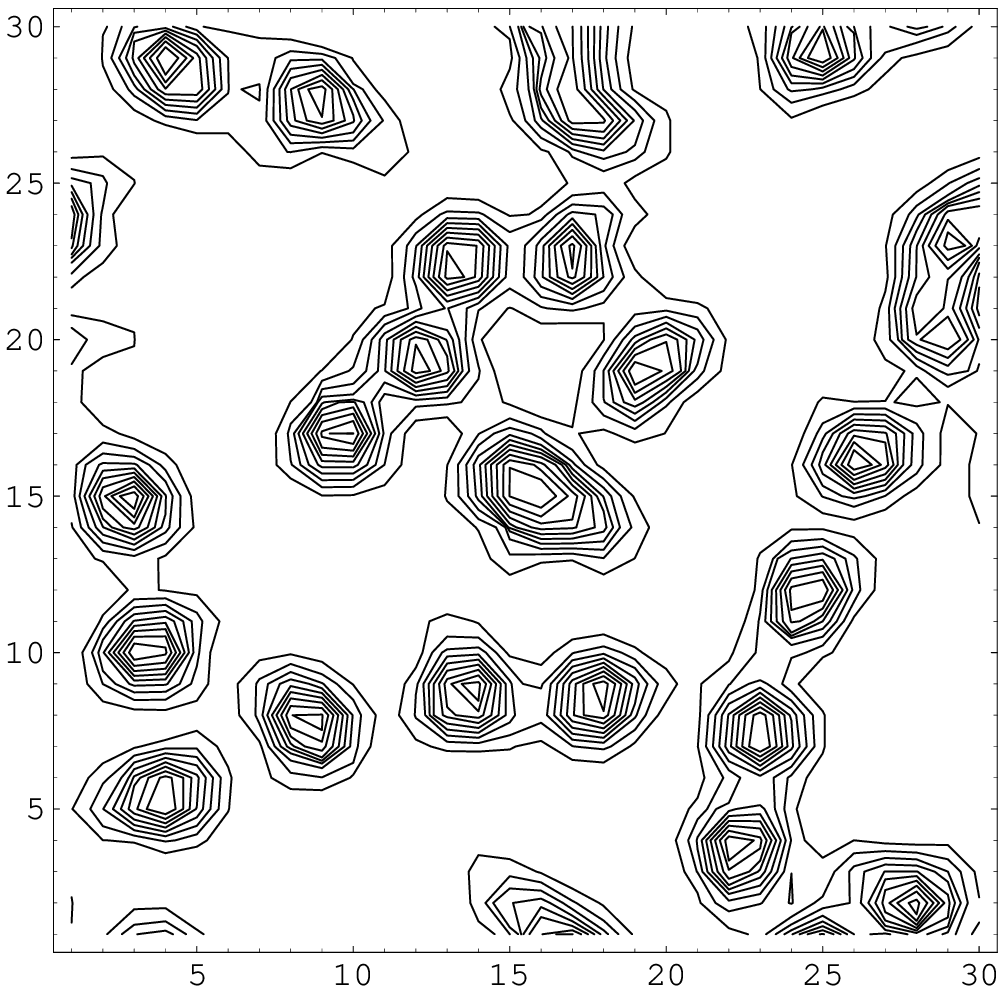}}
%\vspace{-0.5truecm}
\begin{minipage}[t]{8.1cm}
\caption{Contour plots of the charge density in the presence of 
a  strong the tendency towards crystallization for  $\gamma=0.6$.
at $\nu=1/7$ with $N_D=1$ and $N_\phi=175$.
Plot (a) is without the ion while plot (b) is with the ion.
Comparing these two plots we can infer that the ion captures an electron
while the location of the other electrons are unchanged.  The unit
of length is $\ell$.}
\label{figure4}
\end{minipage}
\end{figure}

\end{multicols}


\begin{references}

\bibitem{Kli}K. von Klitzing, G. Dorda, and M. Pepper,  Phys. Rev. Lett. {\bf 45}, 494 (1980).
\bibitem{Lee}D.-H. Lee and Z. Wang, Phys. Rev. Lett. {\bf 76}, 4014 (1996).
\bibitem{Chan}Chan, Glicofridis, Ashoori, and Melloch, Phys. Rev. Lett. {\bf 796}, 2867 (1997).
\bibitem{Yang1}S.-R. Eric Yang and A. H. MacDonald, Phys. Rev. Lett. {\bf 70}, 4110 (1993).
\bibitem{Wang}Z. Wang and Xiong, Phys. Rev. Lett. {\bf 83}, 828 (1999).
\bibitem{Yang2}S.-R. Eric Yang, A. H. MacDonald, and B. Huckestein, 
Phys. Rev. Lett. {\bf 74}, 3229 (1995).
\bibitem{Aoki}H. Aoki and T. Ando, Phys. Rev. Lett. {\bf 54}, 831 (1985).
\bibitem{Cha}J.T. Chalker and G. J. Daniell,  Phys. Rev. Lett. {\bf 61}, 593(1988).
\bibitem{Huo}Y. Huo, R.E. Hetzel, and R. N. Bhatt, Phys. Rev. Lett. {\bf 70}, 481 (1993).
\bibitem{Koch}S.Koch, R.J. Haug, K. von Klitzing, and K. Ploog, Phys. Rev. Lett. {\bf67},
883 (1991).  
\bibitem{Sha}D. Shahar et al., Phys. Rev. Lett. {\bf 7445}, 4511 (1995).
\bibitem{Wangetal} Z. Wang, M.P.P. Fisher, S. Girvin, and J. Chalker,
Phys. Rev. B {\bf61}, 8326 (2000).
\bibitem{Eng}L. W. Engel, D. Shahar, C. Kurdak, and D.C. Tsui,  Phys. Rev. Lett. 
{\bf 71}, 2638 (1993).
\bibitem{Huc}B. Huckestein and M. Backhaus, Phys. Rev. Lett. {\bf 82}, 5100 (1999).
\bibitem{Plee}P.A. Lee, Phys. Rev.B {\bf 26}, 5882 (1982)
\bibitem{atkinson} W. A. Atkinson, P. J. Hirschfeld, and A. H. MacDonald, 
Phys. Rev. Lett. {\bf 85}, 3922 (2000).    
%\bibitem{Mac}W.L. McMillan, Phys. Rev.B, {\bf 24},2739 (1981).
%\bibitem{Imry}Y. Imry, Y. Gefen, and D. J. Bergman, Phys. Rev. B, {\bf 26},3436 (1982).
%\bibitem{Abr}Abrikosov
\bibitem{com} In the calculation of the Thouless number 
by Yang et al\cite{Yang2} 
this self-consistency is kept.
When the Thouless number is calculated from the HF eigenvalues HF orbitals are used in effect.  
They calculated HF orbitals self-consistency {\it after} changing the boundary conditions.
We believe that the 
change in the 
exchange-potential when the boundary conditions are changed is like the ladder sum vertex
corrections.
\bibitem{note} We take $dn/d\mu$ to be defined by the static irreducible density
response function in the long wavelength limit.
In our numerical calculations, we (1) take the d.c. limit,
(2) ensemble average, and (3) take the thermodynamic limit.  
In the metallic state we find that the full screening is achived over
an area that does not diverge with the system size, so that calculating the 
total induced charge is equivalent to taking the limit q to zero after taking
the thermodynamic limit. 
We have used a unit test charge for our calculations because the response to 
fractional charges can be unphysical in the strongly localized limit well away
from half-filling.  Because of this choice we are not strictly 
in the linear response limit, even at half-filling.  However, the response will 
be nonlinear only near the inserted charge.  Since full screening is recovered over
a finite area, we think that taking the linear response limit in the end
would not change our conclusion of a finite $dn/d\mu$ at quantum Hall transition.
Moreover, we expect the total induced charge in the long wavelength
limit to follow linear response.  
\bibitem{Yang3}S.-R. Eric Yang and A. H. MacDonald, Phys. Rev. B {\bf 42}, 10811 (1990).
\bibitem{com1}
It has frequently been argued that $q_{sc}$ vanishes in an Anderson insulator, a view
that is consistent with our numerical result.  See for example, Y. Imry, Y. Gefen, 
and D. J. Bergman, Phys. Rev. B, {\bf 26},3436 (1982).  This opinion appears
not to be universally held, however, and some confusion appears to exist because of the 
glassy nature of localized interacting electron response funstions.
\bibitem{ruzin}I.M. Ruzin, S. Marianer, and B.I. Shklovskii, Phys. Rev. B 
{\bf46 }, 3999(1992).

\end{references}
\end{document}